\documentclass{article}
\usepackage[english]{babel}
\usepackage{mydefs}
\usepackage{todonotes}

\title{Deciding Entailment of Implications with Support and Confidence in Polynomial Space}
\author{
  \small Daniel Borchmann\\
  \small TU Dresden\\
  \small Faculty of Mathematics and Sciences\\
  \small Institute for Algebra\\
  \small \texttt{daniel.borchmann@mailbox.tu-dresden.de}}
\date{\today}

\begin{document}

\maketitle

\begin{abstract}
  Association Rules are a basic concept of data mining.  They are, however, not understood
  as logical objects which can be used for reasoning.  The purpose of this paper is to
  investigate a model based semantic for implications with certain constraints on their
  support and confidence in relational data, which then resemble association rules, and to
  present a possibility to decide entailment for them.
\end{abstract}

\section{Introduction}

Association Rules are a basic pattern in data mining which has undergone extensive
research during the last decades.  Thereby a major focus has been on the question how to
represent the set of all association rules as concisely as possible, as this set might be
too large to be practically helpful.  However, in those considerations representations
based on model semantics have not been used.  Instead, rules based formalisms for infering
association rules from others have been developed~\cite{DBLP:conf/ki/StummeTBPL01,
  DBLP:conf/esf/Kryszkiewicz02}.  One notable exception is the work of
Luxenburger~\cite{Lux1994} who investigated \emph{partial implication}, which are closely
related to association rules and to the constrained implications we shall introduce in
this paper.

In this work we want to investigate this model based approach.  For this, we shall develop
a straight forward model based semantic for \emph{constrained implications}, which shall
replace the original notion of association rules.  The main focus of our work then shall
be the investigation on whether a set of constrained implications \emph{entails} another
constrained implication or not.  We shall show that this question can be rephrased in
terms of two linear programs, which then can be solve in polynomial space.

We shall start our investigation with a short introduction into Formal Concept
Analysis~\cite{fca-book}, which we are going to use as the basic formalism to formulate
our considerations.  After this, we introduce the classical concepts of \emph{support} and
\emph{confidence}.  We then investigate constrained implications and formulate a semantic
based on models for constrained implications.  We shall finally formulate the entailment
problem for constrained implications.

The next step shall then be an equivalent reformulation of a certain instance of the
entailment problem into a set of two linear programs.  The reformulation we are going to
make is quite natural, however the resulting pair of linear programs may be exponentially
large in the size of the original input.  Nevertheless, as we shall show subsequently, we
are able to show that we can solve both linear programs in space polynomial in the size of
the input.  Hence, entailment of constrained implications is decidable in PSpace.

\section{Implications with Support and Confidence}

Let $G$ and $M$ be two sets and let $I\subseteq G\times M$.  We shall call the triple
$\con{K} = (G,M,I)$ a \emph{(formal) context} and shall associate with it the following
interpretation: The elements $g\in G$ are the \emph{objects} of $\con{K}$, the elements
$m\in M$ are the \emph{attributes} of $\con{K}$ and we shall say that \emph{$g$ has the
  attributes $m$} if and only if $(g,m)\in I$.  In that case we may also write $g\relI
m$.  The set of objects of a formal context $\con{K}$ may be denoted $G_\con{K}$, 
likewise $M_\con{K}$ denotes the set of attributes of $\con{K}$ and $I_\con{K}$ denotes
the incidence relation of $\con{K}$.

Let us fix a formal context $\con{K}$ and let $A\subseteq M_\con{K}$.  We shall denote
with
\begin{equation*}
  A' := \set{g\in G \mid \forall m\in A: g\relI m}
\end{equation*}
the set of all common objects of all attributes of $A$.  Likewise, for a set $B\subseteq
G(\con{K})$ the set of common attributes of all objects in $B$ is denoted by
\begin{equation*}
  B' := \set{m\in M \mid \forall g\in B: g\relI m}.
\end{equation*}
The sets $A'$ and $B'$ are called the \emph{derivations} of $A$ and $B$ in $\con{K}$,
respectively, and accordingly the operators $\cdot'$ are the called \emph{derivation
  operators} of $\con{K}$.

Let $A,B\subseteq M_\con{K}$.  We shall call the pair $(A,B)$ an \emph{implication} of
$\con{K}$ and denote it with $A\to B$.  The set of all implications of $\con{K}$ is
denoted by $\Imp(\con{K})$.  We shall say that the implication $A\to B$ \emph{holds} in
$\con{K}$, denoted $\con{K}\models A\to B$, if and only if every object $g\in G_\con{K}$
that has all the attributes of $A$ also has all the attributes of $B$.  Using the
derivation operators, we can comprehensively write this as
\begin{equation*}
  \con{K}\models A\to B \iff A' \subseteq B'.
\end{equation*}

Now let $\con{K}$ be a finite and non-empty formal context, \ie both $G_\con{K}$ and
$M_\con{K}$ are now assumed to be finite and non-empty sets.  The \emph{support of $A$} is
defined to by
\begin{equation*}
  \supp_\con{K}(A) := \frac{|A'|}{|G|}.
\end{equation*}
Similarly, the \emph{support of the implication $A\to B$} is set to be
\begin{equation*}
  \supp_\con{K}(A\to B) := \supp_\con{K}(A).
\end{equation*}
Furthermore, the \emph{confidence} of the implication $A\to B$ is given by
\begin{equation*}
  \conf_\con{K}(A\to B) := \frac{|(A\cup B)'|}{|A'|}
\end{equation*}
if $|A'| \neq 0$.  Otherwise, $\conf_\con{K}(A\to B) := 1$.

We are now ready to introduce the notion of constrained implications.
\begin{Definition}
  Let $\con{K}$ be a finite and non-empty formal context.  Then a \emph{constrained
    implication} of $\con{K}$ is a triple $(A\to B, s, c)$ where $A\to B\in\Imp(\con{K})$
  and $s, c\in[0,1]\cap\QQ$.

  A constrained implication $r = (A\to B, s, c)$ is said to have \emph{minimal support
    $s_0$} and \emph{minimal confidence $c_0$} if $s \ge s_0$ and $c \ge c_0$.  We shall
  denote the set of all constrained implications of $\con{K}$ with minimal support $s_0$
  and minimal confidence $c_0$ by $\conImp_{s,c}(\con{K})$.
\end{Definition}

Constrained implications resemble the notion of association rules, but in a more formal
setting.  In particular, we can ask whether a constrained implication \emph{holds} in an
\emph{arbitrary} formal context or not.
\begin{Definition}
  Let $\con{K}$ be a finite and non-empty formal context and let $r = (A\to B, s, c)$ be
  constrained implication of $\con{K}$.  Then $r$ \emph{holds} in $\con{K}$, written as
  $\con{K} \models r$, if and only if
  \begin{gather*}
    \supp_\con{K}(A\to B) \ge s, \\
    \conf_\con{K}(A\to B) \ge c.
  \end{gather*}
\end{Definition}

Let us fix a set $\mathcal{L} = \set{(A_i\to B_i, s_i, c_i) \mid i\in I}$ of constrainted
implications and let $r = (A\to B, s, c)$ be another constrainted implication.  The
problem we want to investigate in the sequel is the following.
\begin{Definition}[Entailment for Constrainted Implications]~\\
  Given $\mathcal{L}$ and $(A\to B, s, c)$, does then for every formal context $\con{K}$
  hold that
  \begin{equation*}
    \con{K}\models \mathcal{L} \implies \con{K}\models (A\to B, s,c),
  \end{equation*}
  \ie does $\mathcal{L}$ \emph{entail} $(A\to B, s,c)$?
\end{Definition}
We shall write $\mathcal{L}\models (A\to B, s, c)$ if and only if $\mathcal{L}$ entails
$(A\to B, s,c)$.  An \emph{instance} of the entailment problem is just a pair
$(\mathcal{L}, r)$ of a set of constrained implications $\mathcal{L}$ and a single
constrained implication $r$.

\section{Linear Programs for the Entailment Problem}

The purpose of this section is to association with the set $\mathcal{L}$ and the rule $r =
(A\to B, s, c)$ a pair of linear programs such that entailment can be decided by solving
the programs.  Before we do so, however, we shall somehow restrict the sets of possible
formal contexts $\con{K}$ that we have to investigate to decide entailment.

For this we observe that we can restrict the set of attributes of the formal contexts
occurring in the entailment problem to those attributes which actually occur somewhere in
the implications involved.  Define
\begin{equation*}
  M := (A\cup B)\cup \bigcup_{i\in I}A_i\cup B_i.
\end{equation*}
Then the following statement holds:
\begin{Lemma}
  \label{M suffices}
  Let $\mathcal{L}$ and $r = (A\to B, s, c)$ as before.  Then $\mathcal{L} \models (A\to
  B, s, c)$ if and only if for all formal contexts $\con{K}$ with attribute set $M$ it
  holds that
    \begin{equation*}
      \con{K}\models \mathcal{L} \implies \con{K}\models (A\to B, s, c).
    \end{equation*}
  \begin{Proof}
    The direction $(\Rightarrow)$ is clear.  The main idea for the other direction is that
    implications can only ``restrict'' the occurrence of attributes they contain.  More
    formally, let us assume that $\mathcal{L}\not\models r$.  We shall show that then
    there exists a finite and non-empty formal context $\con{K}$ with attributes set $M$
    such that $\con{K}\models \mathcal{L}$ and $\con{K}\not\models r$.  Since $\mathcal{L}
    \not\models r$ there exists a formal context $\con{K}_0$ with $M\subseteq
    M_{\con{K}_0}$ such that $\con{K}_0\models \mathcal{L}$ and $\con{K}_0\not\models r$.
    Let $\con{K}$ be the formal context that arises from $\con{K}_0$ by deleting all
    attributes $M_{\con{K}_0}\setminus M$.  More formally,
    \begin{equation*}
      \con{K} := (G_{\con{K}_0}, M, I_{\con{K}_0}\cap (G_{\con{K}_0}\times M)).
    \end{equation*}
    Now if $X\subseteq M$, then the derivation of $X$ in $\con{K}_0$ and $\con{K}$ is the
    same, since
    \begin{equation*}
      g\relI_{\con{K}_0} m \iff g\relI_\con{K} m
    \end{equation*}
    holds for all $g\in G_{\con{K}_0}$ and $m\in X$.  Therefore, the support and
    confidence of all elements in $\mathcal{L}$ and of the rule $r$ is the same on both
    $\con{K}_0$ and $\con{K}$ and hence $\con{K}\models \mathcal{L}, \con{K}\not\models
    r$ as required.
  \end{Proof}
\end{Lemma}

We are now going to construct the aforementioned linear programs for our instance
$(\mathcal{L}, r)$ of the entailment problem.

Let $\con{K}$ be a formal context with attribute set $M$.  For every set $A\subseteq M$
let $x_A$ be the number of objects $g$ in $\con{K}$ such that $g' = A$, divided by $|G|$, \ie
\begin{equation*}
  x_A := \frac{|\set{g\in G\mid g' = A}|}{|G|}.
\end{equation*}
Now if $r_i = (A_i\to B_i, s_i, c_i)\in\mathcal{L}$ is such that $\con{K}\models r_i$,
then
\begin{align*}
  \supp_\con{K}(A_i\to B_i) &= \supp_\con{K}(A_i) \ge s_i,\\
  \conf_\con{K}(A_i\to B_i) &\ge c_i.
\end{align*}
This can we rewritten with the help of the variables $x_A$ as
\begin{equation}
  \label{basic inequalities}
  \begin{gathered}
    \sum_{A\supseteq A_i} x_A \ge s_i,\\
    \sum_{A\supseteq A_i\cup B_i}x_A - c_i\sum_{A\supseteq A_i}x_A \ge 0.
  \end{gathered}
\end{equation}
The second inequality stems from the fact that $\supp_\con{K}(A_i\cup B_i) -
c_i\supp_\con{K}(A_i) \ge 0$.

Additionally, the variables $x_A$ satisfy the equation $\sum_{A\subseteq M}x_A = 1$, which
can be rewritten as two inequalities:
\begin{equation}
  \label{entries sum to one}
  \begin{aligned}
    \sum_{A\subseteq M}x_A &\ge 1,\\
    -\sum_{A\subseteq M}x_A &\ge -1.
  \end{aligned}
\end{equation}
Finally, all values $x_A$ satisfy $x_A \ge 0$ and $x_A\in\QQ$.

Those inequalities can be rewritten in a compact form using matrices.  Let ${\bf
  A}\in\QQ^{(2|I|+2)\times 2^{|M|}}$ and $b\in\QQ^{2|I|+2}$ be such that the
aforementioned inequalities are represented as
\begin{equation}
  \label{inequality system}
  {\bf A}{\bf x} \ge {\bf b}, \quad {\bf x} \ge {\bf 0}.
\end{equation}
Thereby the vector ${\bf x} = (x_i)_{i=1,\ldots,2^{|M|}}$ shall be such that the variable
$x_i$ corresponds to the variable $x_A$ where the binary representation of $i$ corresponds
to the characteristic vector of $A$ as a subsets of $M$, for some arbitrary but fixed
linear ordering of $M$.

\begin{Example}
  Let us illustrate the construction by means of a simple example.  Let
  \begin{equation*}
    \mathcal{L} = \set{(\set{a}\to \set{b}, \nicefrac{1}{2}, \nicefrac{1}{3}),
                       (\set{a}\to \set{c}, \nicefrac{1}{3}, \nicefrac{1}{4})}.
  \end{equation*}
  Then $M = \set{a,b,c}$.  Let us order $M$ by virtue of $a > b > c$.  Then
  \begin{gather*}
    {\bf A} =
    \begin{pmatrix}
      0&0&0&0&1&1&1&1\\
      0&0&0&0&-\nicefrac{1}{3}&-\nicefrac{1}{3}&1-\nicefrac{1}{3}&1-\nicefrac{1}{3}\\
      0&0&0&0&1&1&1&1\\
      0&-\nicefrac{1}{4}&0&-\nicefrac{1}{4}&0&1-\nicefrac{1}{4}&0&1-\nicefrac{1}{4}\\
      1&1&1&1&1&1&1&1\\
      -1&-1&-1&-1&-1&-1&-1&-1
    \end{pmatrix}
    \intertext{and}
    {\bf b} =
    \begin{pmatrix}
      \nicefrac{1}{2}\\
      0\\
      \nicefrac{1}{3}\\
      0\\
      1\\
      -1
    \end{pmatrix}
  \end{gather*}
  Here we have used the aforementioned convention, namely that the index $i$ corresponds
  to the subsets $A_i$ of $M$ whose characteristic vector with respect to the chosen
  linear ordering of $M$ is the binary representation of $i$.  Hence, $i = 0$ corresponds
  to the empty set, $i = 1$ to the set $\set{c}$ and so on.
\end{Example}

\begin{Theorem}
  \label{solutions are models are solutions}
  Let $\mathcal{L} = \set{(A_i\to B_i, s_i, c_i) \mid i\in I}$ be a set of constrainted
  implications and let $M := \bigcup_{i\in I}A_i\cup B_i$.  Then every context $\con{K}$
  with attribute set $M$ and $\con{K} \models \mathcal{L}$ induces a solution ${\bf
    x}_\con{K}$ of~(\ref{inequality system}).  Conversely, every solution ${\bf x}$
  of~(\ref{inequality system}) induces a nonempty collection $\mathcal{K}$ of formal
  context $\con{K}$ with attribute set $M$, $\con{K}\models \mathcal{L}$ and ${\bf
    x}_\con{K} = {\bf x}$.
\end{Theorem}

We therefore see that there exists a one-to-one correspondence between non-empty classes
of models of $\mathcal{L}$ and the solutions of~(\ref{inequality system}).

\begin{Proof}
  From the previous considerations we have seen that every formal context $\con{K}$ with
  $\con{K} \models \mathcal{L}$ gives rise to a solution of~(\ref{inequality system}).

  For the converse direction let ${\bf x}$ be a solution of~(\ref{inequality system}).
  Then the entries $x_i$ of ${\bf x}$ are rational numbers.  Let $n$ be the least common
  denominator of all entries of $x_i$.  Then define the formal context $\con{K}_{\bf x}$
  with attribute set $M$ as follows: For every index $i\in\set{1,\ldots,2^{|M|}}$ we add
  $x_i\cdot n$ objects to $\con{K}_{\bf x}$ whose intent is the set $A$ that corresponds
  to the binary representation of $i$.  Then $|G_{\con{K}_{\bf x}}| = n$ and we claim that
  $\con{K}_{\bf x}$ is a model of $\mathcal{L}$.

  For this let $i\in I$.  Then
  \begin{equation*}
    \supp_{\con{K}_{\bf x}}(A_i \to B_i) = \sum_{j\mid A_j\supseteq A_i}x_j \ge s_i
  \end{equation*}
  where $A_j$ is the corresponding set for the index $j$.  Likewise it follows
  $\conf_{\con{K}_{\bf x}}(A_i \to B_i) \ge c_i$, because ${\bf x}$ is a solution
  for~(\ref{inequality system}).  From the construction of $\con{K}_{\bf x}$ and the
  construction of ${\bf x}_{\con{K}_{\bf x}}$ it is apparent that ${\bf x} = {\bf
    x}_{\con{K}_{\bf x}}$.
\end{Proof}

The crucial observation now is that we can transform the problem of deciding entailment
into a pair of linear optimization problems.  The key idea is that we can read the
entailment problem $\mathcal{L} \models (A\to B, s, c)$ as the question whether the minimal
possible support for $A\to B$ in every model of $\mathcal{L}$ can be lower then $s$;
similarly, one can ask what the minimal confidence of $A\to B$ in all models of
$\mathcal{L}$ is lower than $c$.  Therefore, entailment is equivalent to solving two linear
optimization problems over $\QQ$.

\begin{Theorem}
  Let $\mathcal{L}\cup\set{(A\to B, s, c)}$ be a set of constrainted implications, $|M| =
  \bigcup_{i\in I}A_i\cup B_i\cup A \cup B$ and let ${\bf A}$ and ${\bf b}$ as defined
  for~(\ref{inequality system}).  Furthermore, let
  \begin{equation*}
    \mathcal{X} := \set{{\bf x}\in\QQ^{2^{|M|}}
      \mid {\bf A}{\bf x} \ge {\bf b}, {\bf x} \ge {\bf 0}}.
  \end{equation*}
  Then $\mathcal{L} \models (A\to B, s, c)$ if and only if
  \begin{equation}
    \label{entailment condition}
    \begin{aligned}
      s &\le \min\set{\sum_{X\supseteq A}x_X \mid {\bf x}\in\mathcal{X}}\\
      0 &\le \min\set{\sum_{X\supseteq A\cup B}x_X - c\sum_{X\supseteq A}x_X \mid {\bf x}
        \in \mathcal{X}}.
    \end{aligned}
  \end{equation}
  \begin{Proof}
    Let $\mathcal{L} \models (A\to B, s, c)$.  Then for every formal context $\con{K}$
    with attribute set $M$ and $\con{K} \models \mathcal{L}$ it holds
    \begin{gather*}
      \supp_\con{K}(A\to B) \ge s,\\
      \conf_\con{K}(A\to B) \ge c,
    \end{gather*}
    and so for the corresponding solution ${\bf x}_\con{K}$ of~(\ref{inequality system})
    we have
    \begin{gather*}
      \sum_{X\supseteq A}x_X \ge s,\\
      \sum_{X\supseteq A\cup B}x_X - c\sum_{X\supseteq A}x_X \ge c.
    \end{gather*}
    Since by Theorem~\ref{solutions are models are solutions} every solution
    of~(\ref{inequality system}) originates from a model $\con{K}$ of $\mathcal{L}$, the
    inequalities~(\ref{entailment condition}) are satisfied.

    Conversely, let $\mathcal{L}\not\models(A\to B, s, c)$.  Then by Lemma~\ref{M
      suffices} there exists a formal context $\con{K}$ with attribute set $M$ such that
    $\con{K}\models \mathcal{L}$ and $\con{K}\not\models(A\to B, s, c)$, \ie
    \begin{gather*}
      \supp_\con{K}(A\to B) < s\text{ or}\\
      \conf_\con{K}(A\to B) < c.
    \end{gather*}
    Then the corresponding solution ${\bf x}_\con{K}$ of~(\ref{inequality system})
    satisfies
    \begin{gather*}
      \sum_{X\supseteq A}x_X < s \text{ or}\\
      \sum_{X\supseteq A\cup B}x_X - c\sum_{X\supseteq A}x_X < c.
    \end{gather*}
    and so the inequalities~(\ref{entailment condition}) are not satisfied.
  \end{Proof}
\end{Theorem}

\section{Deciding Entailment in Polynomial Space}

The main disadvantage of the above approach is the size of the system~(\ref{inequality
  system}), which is exponential in $|M|$ and might therefore be exponential in the size
of the original input $\mathcal{L}$ and $r = (A\to B, s, c)$.  The resulting linear
programs~(\ref{entailment condition}) might therefore have a number of variables which is
exponential in the number of inequalities.  The case, however, that the number of
variables is much larger than the number of restrictions is quite common in linear
programming and algorithms have been devised to handle them specially.

The purpose of this section is to represent such an algorithm.  We shall show that this
algorithm then is able to solve~(\ref{entailment condition}) in space polynomial in the
size of the original input $\mathcal{L}$ and $r$.

To properly formulate this claim we need to agree on an encoding system for $\mathcal{L}$
and $r$, for otherwise the notion of \emph{polynomial space in the size of the input}
would be meaningless. See~\cite{complexity-theory-garey-johnson} for more details.

We shall agree on the convention that integers are stored as binary numbers. Rational
numbers can be stored as pairs of integers.  Matrices and vectors are stored as sequences
of their entries, in row-major order.  Finally, sets are represented as sequences of the
elements they contain.  For convenience we shall assume that we can use special characters
like braces, parentheses and commas to separate different entities.  This only increases
the space requirements by a negligible amount.

As the following description of the algorithm gets technical at certain points we may
adopt some syntactical conventions to make this description more pragmatic.

If ${\bf x}$ is a rational vector, we may write ${\bf x} = ({\bf x}_1, {\bf x}_2)$ to
denote two vectors ${\bf x}_1, {\bf x}_2$ whose entries are a partition of the entries of
${\bf x}$.  That does not necessarily mean that the vector ${\bf x}$ must be ordered such
that ${\bf x}_1$ comes first and ${\bf x}_2$ second.  The same notation may be used to
denote column partitions of matrices, \ie that the columns of two matrices ${\bf A}_1,
{\bf A}_2$ constitute the columns of a matrix ${\bf A}$ may be written as ${\bf A} = ({\bf
  A}_1, {\bf A}_2)$, without any special assumptions on the ordering of any columns.
Finally, the columns of a matrix ${\bf A}$ are denoted by $({\bf a}_i)_{i=1,\ldots,n}$, if
${\bf A}$ has $n$ columns.

\subsection{The Revised Simplex Method}

One of the best known algorithms for solving linear programs is the \emph{Simplex Method}
devised by George Dantzig.  The \emph{revised Simplex Method} is a variation of the
original Simplex Method that reduces the space needed for the actual computations.  The
revised Simplex Method is mathematical folklore and we shall only give a short and compact
representation of the method without giving any proofs.  For further details, we refer to
any introductory text on linear programming.  The following description
follows~\cite{ADMII-Groetschel, Numerik-Grossmann-Terno}.

Let us consider the general linear optimization problem
\begin{equation}
  \label{general lp}
  \max\set{{\bf c}^T{\bf x}
    \mid {\bf x}\in\QQ^n, {\bf Ax} = {\bf b}, {\bf x} \ge {\bf 0}}
\end{equation}
where ${\bf A}\in\QQ^{m\times n}$ and ${\bf b}\in\QQ^m$.  Moreover, we assume $m \le n$
and that ${\bf A}$ has maximal rank, \ie $\rank({\bf A}) = m$.  Then the vertexes of the
polytope\footnote{A \emph{polytope} is finite intersection of halfspaces in $\QQ^n$.} $P =
\set{{\bf x} \mid {\bf Ax} = {\bf b}}$ contain optimal solutions for~(\ref{general lp}),
and the Simplex Method (as well as the revised Simplex Method) goes along the edges of the
polytope to find an optimal vertex.  Vertexes of $P$ correspond to maximal independent
sets $B = \set{{\bf a}_{i_1}, \ldots, {\bf a}_{i_k}}$ of certain columns of ${\bf A}$, and
since ${\bf A}$ has full rank, $k = m$ and the set $B$ can be understood as a regular
submatrix ${\bf B}$ of ${\bf A}$. Let us denote with ${\bf N}$ the submatrix of ${\bf A}$
that contains all columns except those in $B$.

Starting with such a submatrix ${\bf B}$, which corresponds to a vertex ${\bf x}$ of $P$,
the revised Simplex Method now tries to find another vertex ${\bf x}'$ such that ${\bf
  c}^T{\bf x} \le {\bf c}^T{\bf x}'$.  If such a vertex exists, then there also exists a
vertex which is linked to ${\bf x}$ by an edge of the polytope $P$.  To find such a vertex
${\bf x}'$, which amounts to find a set $B'$ or the corresponding submatrix ${\bf B'}$ of
${\bf A}$, the revised Simplex Method proceeds as follows:
\begin{enumerate}[i. ]
\item\label{rsm: init} Subdivide the vectors ${\bf x} = ({\bf x}_B, {\bf x}_N)$ and ${\bf
    c} = ({\bf c}_B, {\bf c}_N)$ according to the chosen columns of ${\bf A}$ in $B$.
  Then ${\bf x}_N = {\bf 0}$ and ${\bf x}_B = {\bf B}^{-1}{\bf b}$.
\item\label{rsm:compute r} Compute ${\bf r} = {\bf c}_N^T - {\bf c}_B^T{\bf B}^{-1}{\bf
    N}$.
\item\label{rsm:optimality check} If all entries in ${\bf r}$ are non-positive, then
  ${\bf x}$ is an optimal solution for~(\ref{general lp}).
\item\label{rsm:choose direction} Otherwise let $k$ be such that the $k$th entry in ${\bf
    r}$ is negative.
\item\label{rsm:boundedness check} If all entries in the $k$th column of ${\bf B}^{-1}{\bf
    N}$ are non-positive, then~(\ref{general lp}) does not have a solution (the value of
  ${\bf c}^T{\bf x}$ is unbounded over $P$).  Abort.
\item\label{rsm:choose base element} Otherwise let $s$ be an index such that
  $\frac{x_s}{d_{sk}}$ is minimal with $d_{sk} > 0$.  Here $x_j$ is the $j$th component of
  ${\bf x}_B$ and the entries of ${\bf B}^{-1}{\bf N}$ are the elements $d_{ij}$.  Thereby
  choose $s$ such that the corresponding column ${\bf a}_s$ of ${\bf A}$ is the
  lexicographically smallest column such that $\frac{x_s}{d_{sk}}$ is minimal.
\item\label{rsm:update} Now the $s$th column ${\bf a}_s$ of ${\bf A}$ is an element of $B$
  and the $k$th column ${\bf a}_k$ of ${\bf A}$ is not an element of $B$.  Define $B' :=
  B\setminus \set{{\bf a}_s}\cup\set{{\bf a}_k}$.
\end{enumerate}

It can be shown that the solution ${\bf x}'$ corresponding to the set $B'$ is a vertex of
$P$ such that ${\bf c}^T{\bf x}' \ge {\bf c}^T{\bf x}$.  Iterating this procedure until it
either returns an optimal solution or the information that no optimal solution exists
constitutes the optimization procedure of the revised Simplex Algorithm.  By (10.6)
of~\cite{ADMII-Groetschel}, this algorithm always terminates.

Now that we know the necessary details of the revised Simplex Method, we are able to prove
the following theorem.

\begin{Theorem}
  \label{rsm}
  Let ${\bf A}\in\QQ^{m\times n}, {\bf b}\in\QQ^m, {\bf c}, {\bf x}'\in\QQ^n$ such that
  $m\le n$, $\rank({\bf A}) = m$ and ${\bf Ax'} = {\bf b}$.  Let $k$ be the largest amount
  of space needed to store any entry of ${\bf A}, {\bf b}, {\bf c}$.  Then the linear
  program
  \begin{equation*}
    \max\set{{\bf c}^T{\bf x}
      \mid {\bf x}\in\QQ^n, {\bf Ax} = {\bf b}, {\bf x} \ge {\bf 0}}
  \end{equation*}
  can be solved in space $\bigO(km^3)$ in addition to the space needed to represent the
  matrix ${\bf A}$ and the vectors ${\bf b}$ and ${\bf c}$.
  \begin{Proof}
    To see that the statement holds we have to show that we can execute all steps of the
    revised Simplex Method in not more that $\bigO(km^3)$ space or that we can replace
    them by equivalent computations which can be done in space $\bigO(km^3)$.

    We start by observing that we only need to store the indexes of the elements of $B$,
    ${\bf x}_B$ and ${\bf c}_B$.  This can be done in space $\bigO(km)$.  Computing the
    inverse can be done in space $\bigO(km^3)$, using the adjugate matrix of ${\bf B}$.
    This involves computations of determinants of ${\bf B}$ and of submatrices of ${\bf
      B}$ with one less row and column.  Those computations can be carried out in time
    $\bigO(km^3)$ each and hence do not need more space than $\bigO(km^3)$.

    The steps (\ref{rsm:compute r})--(\ref{rsm:choose direction}) can be replaced by a
    simple loop that computes every component of ${\bf r}$ separately and returns the
    index of the first positive component.  For this we note that we only need to compute
    ${\bf c}_B^T{\bf B}^{-1}$ once, yielding a row vector.  After this, we successively
    compute scalar products with rows of ${\bf N}$ and save the index of the first row
    where the result is negative.  If there is no such index, the algorithms stops.  All
    this can be done in space $\bigO(km)$.

    It is clear that we can execute step~(\ref{rsm:boundedness check}) in linear space,
    since the columns of ${\bf B}^{-1}{\bf N}$ have $m$ entries.

    Step~(\ref{rsm:choose base element}) iterates through all the indexes $s$ of columns
    ${\bf a}_s$ in ${\bf B}$.  For those indexes the value $\frac{x_s}{d_{sk}}$ needs to
    be checked for being negative, which can be done without actually computing the
    fraction.  Moreover, comparing one such value with another can also be done without
    computing the value of the fraction.  Finally, the column ${\bf a}_s$ may have to be
    compared to another column vector for being lexicographically smaller.  This can be
    done in space $\bigO(km)$ and hence the overall step can be executed in space
    $\bigO(km)$.

    Finally, we note that updating the value for $B$ requires only space logarithmic in
    $m$.

    So overall, one optimization step of the revised Simplex Method requires not more than
    $\bigO(km^3)$ of space in addition to the space needed to store the original input.
  \end{Proof}
\end{Theorem}

It may be noted that the revised Simplex Method (and the Simplex Method as well) consists
of two phases, from which we only have described the second.  Since we suppose a starting
solution ${\bf x}$, the description as it stands might not be very helpful in practice,
since such a solution might not be given (and might not even exist).  However, in our
special circumstances we can explicitly give such a solution, as we shall see in the next
subsection.

\subsection{Deciding Entailment in Polynomial Space}
\label{subsec:solving the system}

With the revised Simplex Method at hand we are able to show that the linear
programs~(\ref{entailment condition}) can be solved in polynomial space.  The main obstacle
is to transform~(\ref{entailment condition}) into the form of~(\ref{general lp}).

Let us consider the system~(\ref{inequality system}).  We can convert it into the form
of~(\ref{general lp}) by adding \emph{slack variables} $y_i$ for each inequality
in~(\ref{basic inequalities}).  More precisely, we can transform the
inequalities~(\ref{basic inequalities}) into the equations
\begin{gather*}
  \sum_{A\supseteq A_i} x_A - y_{i,1} = s_i,\\
  \sum_{A\supseteq A_i\cup B_i}x_A - c_i\sum_{A\supseteq A_i}x_A - y_{i,2} = 0,
\end{gather*}
such that $y_{i,j} \ge 0$ for all $i\in\set{1,\ldots,|I|}, j\in\set{1,2}$.  We can do
likewise for the inequalities in~(\ref{entries sum to one}) by introducing $y_{1}$
and $y_{2}$.  Hence we obtain the system
\begin{equation}
  \label{transformed inequality system}
  {\bf Ax} - {\bf y} = {\bf b}, \quad {\bf x}, {\bf y} \ge {\bf 0}
\end{equation}
where ${\bf y} = ((y_{i,j})_{i\in I, j\in\set{0,1}}, y_1, y_2)$.  The following
observation is well known in the theory of linear programming.

\begin{Lemma}
  If ${\bf x}$ is a solution of~(\ref{inequality system}), then $({\bf x}, {\bf
    y})\in\QQ^{2^{|I|} + 2|I|+2}$ is a solution of~(\ref{transformed inequality system})
  where ${\bf y} = {\bf Ax} - {\bf b} \ge {\bf 0}$.  Conversely, if $({\bf x}, {\bf y})$
  is a solution of~(\ref{transformed inequality system}), then ${\bf x}$ is a solution
  of~(\ref{inequality system}).

  In particular, the linear programs~(\ref{entailment condition}) and the corresponding
  programs
  \begin{equation}
    \label{transformed entailment condition}
    \begin{gathered}
      \min\set{\sum_{X\supseteq A}x_X \mid {\bf x}\in\mathcal{X}}, \\
      \min\set{\sum_{X\supseteq A\cup B}x_X - c\sum_{X\supseteq A}x_X \mid {\bf x}
        \in \mathcal{X}}
    \end{gathered}
  \end{equation}
  with
  \begin{equation*}
    \mathcal{X} = \set{{\bf x} \mid {\bf Ax} - {\bf y} = {\bf b}, {\bf x, y} \ge {\bf 0}}
  \end{equation*}
  have the same values, respectively.
\end{Lemma}

We are now ready to formulate the overall goal of our considerations.

\begin{Theorem}
  \label{solvable in PSpace}
  The systems~(\ref{transformed entailment condition}) can be solved in polynomial space.
\end{Theorem}

\begin{Corollary}
  The linear programs~(\ref{entailment condition}) can be solved in polynomial space.
  Therefore, entailment of constrained implications can be decided in polynomial space.
\end{Corollary}

\begin{Proof}[Theorem~\ref{solvable in PSpace}]
  Let $m = |I|$.  Let us write the system~(\ref{transformed inequality system}) as
  \begin{equation}
    \label{compact transformed inequality system}
    {\bf \tilde A \tilde x} = {\bf b}, \quad {\bf \tilde x} \ge {\bf 0}
  \end{equation}
  where ${\bf \tilde A} = ({\bf A}, {\bf -I})\in\QQ^{(2m+2)\times(2^m + 2m + 2)}$ and
  ${\bf \tilde x} = ({\bf x}, {\bf y})\in\QQ^{2^m + 2m + 2}$.  Here, ${\bf I}$ denotes the
  identity matrix of the corresponding size.  Note that ${\bf \tilde A}$ has full rank,
  since it contains ${\bf -I}$ as submatrix.

  We start by noting the following two facts, which we shall discuss in detail afterwards:
  \begin{enumerate}[i. ]
  \item\label{idea:implicit representation} The matrix ${\bf\tilde A}$ can be represented
    implicitly, \ie the entries of ${\bf \tilde A}$ can directly be inferred from the
    input $\mathcal{L}$ and $r$.
  \item\label{idea:explicit solution} A solution for~(\ref{compact transformed inequality
      system}) can be given explicitly.  This is due to the fact that we can find an
    explicit solution for~(\ref{inequality system}), namely $x_M = 1$ and $x_A = 0$ for
    $A\subsetneq M$.
  \end{enumerate}

  We start with point~(\ref{idea:implicit representation}).  For this let
  $i\in\set{1,\ldots,m+2}$, $j\in\set{1,\ldots, 2^m + 2m + 2}$ and ${\bf \tilde A} =
  (\tilde a_{st})_{s,t}$.  Then we can distinguish the following cases:
  \begin{enumerate}[1. ]
  \item If $j > 2^m$, then if $i = j - 2^m$, then $\tilde a_{ij} = 1$, otherwise $\tilde
    a_{ij} = 0$.
  \item $i = 2m + 1$ and $j \le 2^m$, then $\tilde a_{ij} = 1$.
  \item $i = 2m + 2$ and $j \le 2^m$, then $\tilde a_{ij} = -1$.
  \item $i = 2k$ for some $k\in I$, $j \le 2^m$.  Let $C$ be the set corresponding to the
    index $j$.  Then $\tilde a_{ij} = 1$ if $C\supseteq A_k$ and $\tilde a_{ij} = 0$
    otherwise.
  \item $i = 2k + 1$ for some $k\in I$, $j \le 2^m$.  Again let $C$ be the set
    corresponding to the index $j$.  If $C\supseteq A_k\cup B_k$, then $C\supseteq A_k$
    and $\tilde a_{ij} = 1 - c_k$.  If $C\not\supseteq A_k\cup B_k$, but $C\supseteq A_k$,
    then $\tilde a_{ij} = -c_k$.  Otherwise, $\tilde a_{ij} = 0$.
  \end{enumerate}
  The last two cases can be seen from the inequalities~(\ref{basic inequalities}).  Hence,
  ${\bf \tilde A}$ can be inferred from $\mathcal{L}$ and $r$ and need not be stored
  explicitly.

  For the point~(\ref{idea:explicit solution}) let ${\bf x}$ as discussed there.  Then let
  ${\bf y} = {\bf Ax - b}$.  Then $B = \set{{\bf a}_{2^m}, {\bf a}_{2^m+1}, \ldots, {\bf
      a}_{2^m + 2m + 1}}$ is a set of linearly independent columns of ${\bf A}$ that
  correspond to the solution $({\bf x}, {\bf y})$.

  Now we can apply Theorem~\ref{rsm} to see that we only need $\bigO(km^3)$ additional
  space to solve~(\ref{compact transformed inequality system}), where $k$ is the largest
  amount needed to store an entry of ${\bf A}, {\bf b}, {\bf c}$.  This however reduces to
  the largest amount needed to store any value of $s_i$ or $c_i$, which are part of the
  input.  Since we have not used more than $\bigO(km^3)$ space so far, we can hence
  solve~(\ref{compact transformed inequality system}) in polynomial space in the size of
  the input and the claim is proven.
\end{Proof}

\section{Conclusions and Further Research}

Starting from the motivation to understand association rules as logical objects usable for
reasoning, we have introduced the notion of constrained implications.  Based on this
definition we have developed semantics for constrained implications based on models and
stated the corresponding entailment problem.  We then investigated a reformulation of
instances of the entailment problem as pairs of linear programs and were able to show that
those linear programs can be solved in polynomial space.

The result that entailment for constrained implications can be decided in polynomial space
does not seem that pleasing.  Above all, it does not give reasonable constrains on the
time needed to decide entailment, a fact that might be of practical interest.

A better complexity bound might hence be desirable.  It is quite unclear whether one can
expect the entailment problem to be an element of either $\mathcal{NP}$, co-$\mathcal{NP}$
or even $\mathcal{P}$.  It might, however, be worth looking into complexity classes of the
polynomial hierarchy whether the entailment problem is located there.

Another problem, which is closely connected to our original motivation but which has not
been addressed in this work is the following:  Since we have established a notion of
entailment for constrained implications it might be worth searching for \emph{minimal
  non-redundant} sets of constrained implications representing $\conImp_{s,c}(\con{K})$
for a formal context $\con{K}$ and $s,c\in[0,1]\cap\QQ$.  Results in this direction might
give more insight in the usefulness of regarding association rules as logical objects.


\bibliography{general,fca}
\bibliographystyle{plain}

\end{document}